RESEARCH ARTICLE

# Newborn skin reflection: Proof of concept for a new approach for predicting gestational age at birth. A cross-sectional study

Zilma Silveira Nogueira Reis[1,2,3☯¤]*, Gabriela Luiza Nogueira Vitral[2¤‡], Ingrid Michelle Fonseca de Souza[1¤‡], Maria Albertina Santiago Rego[4¤‡], Rodney Nascimento Guimaraes[2☯¤]

1 Center of Health Informatics, Faculty of Medicine, Universidade Federal de Minas Gerais, Belo Horizonte, Minas Gerais, Brazil, 2 Postgraduation Program of Women's Health, Faculty of Medicine, Universidade Federal de Minas Gerais, Belo Horizonte, Minas Gerais, Brazil, 3 Department of Gynecology and Obstetrics, Faculty of Medicine, Universidade Federal de Minas Gerais, Belo Horizonte, Minas Gerais, Brazil, 4 Department of Pediatrics, Faculty of Medicine, Universidade Federal de Minas Gerais, Belo Horizonte, Minas Gerais, Brazil

☯ These authors contributed equally to this work.
¤ Current address: Faculty of Medicine, Universidade Federal de Minas Gerais, Avenida Professor Alfredo Balena, Belo Horizonte, Minas Gerais, Brazil
‡ These authors also contributed equally to this work.
* zilma@medicina.ufmg.br

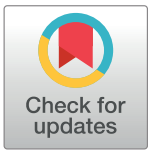

**Data Availability Statement:** All relevant data are within the paper and its Supporting Information files.

**Funding:** This research was supported by the Grand Challenges Exploration from the Bill & Melinda Gates Foundation (Grant number OPP1128907, Contract), http://www.gatesfoundation.org/, and Fundação de Amparo a Pesquisa de Minas Gerais, Brazil, http://www.fapemig.br/en/. Revision in manuscript written in English was supported by Pro-Reitoria de Pesquisa

## Abstract

### Background

Current methods to assess the gestational age during prenatal care or at birth are a global challenge. Disadvantages, such as low accessibility, high costs, and imprecision of clinical tests and ultrasonography measurements, may compromise health decisions at birth, based on the gestational age. Newborns' organs and tissues can indirectly indicate their physical maturity, and we hypothesized that evolutionary changes in their skin, detected using an optoelectronic device meter, may aid in estimating the gestational age. This study analyzed the feasibility of using newborn skin reflectance to estimate the gestational age at birth noninvasively.

### Methods and findings

A cross-sectional study evaluated the skin reflectance of selected infants, preferably premature, at birth. The first-trimester ultrasound was the reference for gestational age. A prototype of a new noninvasive optoelectronic device measured the backscattering of light from the skin, using a light emitting diode at wavelengths of 470 nm, 575 nm, and 630 nm. Univariate and multivariate regression analysis models were employed to predict gestational age, combining skin reflectance with clinical variables for gestational age estimation. The gestational age at birth of 115 newborns from 24.1 to 41.8 weeks of gestation correlated with the light at 630 nm wavelength reflectance 3.3 mm/6.5 mm ratio distant of the sensor, at the forearm and sole (Pearson's correlation = 0.505, P < 0.001 and 0.710, P < 0.001, respectively). The best-combined variables to predict the gold standard gestational age at birth of Universidade Federal de Minas Gerais, Brazil PRPq Call number 02/2017, http://www.ufmg.br/prpq. The funders had no role in study design, data collection and analysis, decision to publish, or preparation of the manuscript.

**Competing interests:** These authors declare a patent deposit number BR1020160256020, on behalf of the Universidade Federal de Minas Gerais and Fundação de Amparo a Pesquisa de Minas Gerais. The inventors were Zilma Silveira Nogueira Reis and Rodney Nascimento Guimaraes. This work was supported by the Bill & Melinda Gates Foundation, Grant Number OPP1128907 Contract. This does not alter our adherence to ONE policies on sharing data and materials.

**Abbreviations:** CCI, intraclass correlation coefficient; CI, confidence interval; GA, gestational age; LED, light emitting diode; LMP, last menstrual period; NICU, neonatal intensive care unit; R, skin reflectance (the measure of skin reflection); SD, Standard deviation.

was the skin reflectance at wavelengths of 630 nm and 470 nm in combination with birth weight, phototherapy, and adjusted to include incubator stay, and sex ($R^2$ = 0.828, P < 0.001). The main limitation of the study is that it was very specific to the premature population we studied and needs to be studied in a broader spectrum of newborns.

## Conclusions

A novel automated skin reflectometer device, in combination with clinical variables, was able to predict the gestational age and could be useful when the information is in doubt or is unknown. Multivariable predictive models associated the skin reflectance with easy to obtain clinical parameters, at the birth scenario. External validation needs to be proven in an actual population with the real incidence of premature infants.

## Introduction

Timely decisions about the immediate care at birth often depend on the gestational age (GA). Perinatal morbidity and mortality are frequently associated with short gestation at birth and with low birth weight in pregnancy [1]. Premature newborns are more likely to die during the first hours of life or to develop lifelong complications [2]. These infants need critical attention to survive, and their age is one of the primary predictors of neonatal outcomes [3]. The current methods to assess the GA during prenatal care or at birth have disadvantages, such as low accessibility, high costs, and imprecision of results [4–6].

Theoretically, fetal age begins at conception, but this information is difficult to determine accurately. Unknown or inaccurate last menstrual period (LMP) dates result in misclassification of newborns at birth, impacting the proportions of preterm and post-term groups, and resulting in inexact proportions of small for gestational age infants and large for gestational age infants [5, 7, 8]. The gold standard for determining GA is the early obstetric ultrasound assessment that establishes or confirms the number of weeks of gestation during the first trimester [4].

While the GA estimation by current approaches faces challenges, fetal maturity may be indirectly determined based on their organs and tissues. Evolutionary changes in the skin of neonates contribute to the maturity scores at birth, together with other external and neuro-muscular indicators [9, 10]. The connection of an age-related morphologic pattern of the fetal skin allowed determination of fetal age in human postmortem examination, with high concordance with LMP and early ultrasound [11]. The skin development during intrauterine life is a continuum that involves the juxtaposition and interaction of mesodermal and ectodermal tissues to form a protective barrier, as expected in term neonates [12]. The period of 22 to 40 weeks of gestation correlates with the maturation of the stratum corneum, the biggest protection against dehydration, heat loss, and injury, and is essential for the health of the preterm infant [13].

The corneum stratum, as well other superficial layers of the skin, can be penetrated by light through its thickness and all its components. Accordingly, the surface structure and tissue composition can be noninvasively accessed by light [14]. Optoelectronic systems can obtain backscattered light signals, captured on a photo-detector, and estimate skin fat thickness measurements [15]. Certain wavelengths of the electromagnetic spectrum have the potential to contribute to the prediction of skin thickness and other skin properties. The skin reflectance values, obtained using spectrophotometry, were different between black and white skin, and female and male newborns [16]. Reflectance at 837 nm increased exponentially with the gestational age, independent of race or sex [17].

Based on the rationale of the skin development during fetal life, this study tested the feasibility of using the neonatal skin reflectance to noninvasively estimate the GA at birth with an optoelectronic device, combined with clinical data, in a multivariable predictive model. We hypothesized that light scattering by the skin mainly occurs because of the presence of keratin and collagen in the superficial layers of the skin, which is still developing during the gestational period.

## Methods and materials

### Environment and subjects

This cross-sectional study evaluated a selection of infants that were born from August 2016 to June 2017, in two Brazilian maternity hospitals, according to inclusion and exclusion criteria. The protocol of research was approved by the institutional review boards in Brazil, the Ethics Committees of the Universidade Federal de Minas Gerais, with a national register number in Plataforma Brasil: CAAE 49798915.2.0000.5149. A written informed consent was obtained from each mother on behalf the newborns. The actual birth and neonatal care took place in the university hospital of Universidade Federal de Minas Gerais and the Hospital Sofia Feldman, a public health institution. Both are referral settings for high-risk pregnancies in the perinatal network of the city, Belo Horizonte, Brazil.

The eligibility criteria for subject selection were alive infants who were born with gestational age above 25 weeks of gestation with gestational age calculated using early ultrasound, done before 14 weeks, which is the gold standard for aging based on ultrasound [4]. Fetal diseases that can affect the skin structure, such as fetal hydrops, genodermatoses, absolute oligohydramnios, as observed in Potter sequence, or clinical evidence of intrauterine infections, were not included. In order to obtain a continuum of the skin response along the GA, premature infants were the primary targets during the selection, even if the sample does not represent the real incidence of birth in each week of gestation. The study size was planned based on a previous moderate correlation between the gestational age and the skin reflection [17].

The newborn assessment occurred twice during the first 48 hours after delivery. The first one occurred as soon as possible after birth and the second one was approximately 24 hours from the first. The skin reflectance was measured in two places on each infant's body, over the anterior distal forearm, and on the sole of the foot. This choice attended the patient security recommendation for minimum manipulation of very-low-birth newborns in NICU. The live newborns were evaluated in the nursing care unit, either on the mother's lap or inside an incubator. In the neonatal intensive care unit (NICU), the assessment occurred inside their incubators or in an open heating crib, when that is where they were being cared for, in order to ensure minimum handling and stable clinical conditions. The phototherapy, when in progress, was turned off during the assessment. Obstetric and clinical variables of the neonates were recorded. The Fenton growth chart for preterm infants was the standard for nutritional classification at birth, based on the birth weight [18].

### The optical prototype and the experimental protocol

A noninvasive, handheld, and low cost optoelectronic device was developed to measure the backscattered light signal from the skin, according to the skin thickness and the composition of the tissue. University bioengineering, physics and electronic technician in the Health Informatics Center constructed the prototype. The expectation was that the backscattered light could exhibit modifications according to the skin growing, mirroring the gestational evolution. The prototype is composed of a sensor probe, unit controller, and power supply. The sensor probe is comprised of two blue, two green, and two red light emitting diodes (LEDs) at wavelengths of 470 nm, 575 nm, and 630 nm, a photodiode, a printed circuit board, and an optical barrier that surrounds the photodiode (Fig 1). A sensor converts the light skin response into a frequency that is directly proportional to the amount of reflected light. The LEDs of the same color were positioned side by side at different distances from the photodiode, resulting in six frequency acquisitions: red, green, and blue, positioned at approximately 3.3 mm and 6.5 mm from the emitter. The unit controller recorded eight automated values: the ambient light, the dark current, and one measurement from each LED. The data acquisitions occurred automatically, when the sensor touched the skin, without operator influence, and recorded one time per newborn and location on the body. A cover for the probe, which was easily disinfected, was developed, using a three dimensional printer, and proper user ergonomics, to avoid intense pressure from the operator against the newborn's skin, to keep the reflectivity measurements stable. Details of the sensor design, the signal processor, and the following process of gestational age estimation were patented under number BR1020160256020 on behalf of the Universidade Federal de Minas Gerais, Brazil and their inventors [19].

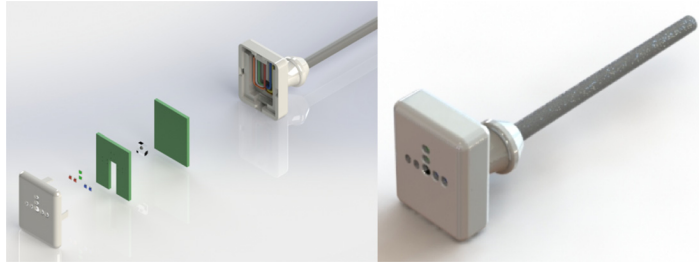

**Fig 1. Prototype of the sensor probe.** (A) Exploded perspective drawing of optoelectronic components fitting in the cover probe. (B) 3D illustration of the second version of the sensor module. Note: Exact positions between the LEDs and photodiode, in the middle of the sensor are: Red 1 = 3.3 mm; Red 2 = 6.5 mm; Green 1 = 4.0 mm; Green 2 = 7.2 mm; Blue 1 = 3.3 mm; Blue 2 = 6.5 mm.

## Data processing and statistical analyses

The descriptive statistics explored the demographic and clinical characteristics of the infants, using frequency, measures of central tendency, and variability, according to the groups of interest, premature and term infants. To evaluate intraobserver variability, additional 122 measurements were obtained in 61 adults. The skin over hand was evaluated, twice and sequentially, by the same observer, with the observer blinded to the previous measurement. One hundred twenty-two measurements were obtained in the same group of 61 adults to assess interobserver variability. The skin over hand was evaluated by a second blinded examiner, with a single measurement being acquired by each one. Intraobserver and interobserver repeatability were assessed using the intraclass correlation coefficient (ICC). Bland–Altman plots [20] were obtained to evaluate the systematic bias between the two measurements for skin reflectance at 470 nm, 575 nm and 630 nm acquisitions, distant near and far from the sensor.

Inferential statistical analysis evaluated the correlation between the skin reflectance and the GA. The sensor acquisitions produced by the skin reflection at 470 nm ($R_{470}$), 575 nm ($R_{575}$), and 630 nm ($R_{630}$), at 3.3 mm and 6.5 mm of distance of the sensor, and their ratios within the same color, were the independent variables. Influence of sex, phototherapy, and incubator environment on the skin reflection ratio was analyzed using a Student t-Test. Paired samples Student-test compared repeated measurements between the first and second day after birth. The best body site on the newborn skin was inferred from the regression coefficients obtained in the graph frequency versus GA. Univariate and multivariate models of regression analysis were employed to estimate the correlation of predictors with the GA. Pearson correlation was the first step, but other non-linear models were adjusted to fit the correlation between predictors and outcome better. Multiple regression analysis included predictor variables from the univariate models, considering the effect modifiers from sex, birth weight, incubator staying, and phototherapy (input and output, P-value of 0.10), using the backward method of model arrangement. The fit of the models and calibration, specifically the ANOVA and Durbin-Watson test for residuals, were performed. Coefficients of determination (adjusted $R^2$) were carried out based on the hypothesis that all coefficients were 0. The statistical program SPSS[R] 22.0 was used for the analysis. The significance levels, adjusted for the hypothesis test, were 5% and 95% (Confidence Interval (95% CI)).

## Results

One, out of the 117 examined newborns, was excluded because the parents withdrew the voluntary authorization for participation. The other exclusion was an outlier for all measurements considered without recuperation. The dependent variable, GA at birth, ranged from 24.1 to 41.8 weeks of gestation. The distribution of their values had a normal standard distribution, with a mean of 34.1 ± 4.1 weeks (95% CI, 33.3 to 34.8) (Fig 2). Most of the subjects had neonatal complications related to high-risk pregnancy and 78% of them were premature (Table 1).

### The newborn skin reflection acquisitions

Regarding the acquisitions, access to the skin reflection from the sole was possible in all of the newborns. In the first 24 hours of life, 113 (98.3%) of the neonates had the forearm region available, depending on which interventions, such as equipment, catheters, and medical devices, were being used for the NICU newborns. Most (76.5%, n = 88) of them had sensors and vascular access at the time of the examination. During the second day of life, 107 (96.8%) and 109 (94.9%) had both the sole and forearm reachable, respectively.

The skin reflection intensity of the skin over the forearm and from the sole, the backscattering dropped on the second day for the $R_{630}$ at 6.5 mm, and for the $R_{575}$ at 4.0 mm and $R_{470}$ at 3.3 mm. Forearm skin reflections were different from the ones obtained at the sole position (Table 2).

### Repeatability of acquisitions

The ICC of differences between repeat acquisitions by the same observer was 0.964 (95% CI, 0.941–0.978) for $R_{630}$ at 3.3 mm, 0.986 (95% CI, 0.977–0.988) for $R_{630}$ at 6.5 mm, 0.996 (95% CI, 0.994–0.998) for $R_{575}$ at 7.2 mm, 0.992 (95% CI, 0.987–0.995) for $R_{575}$ at 4.0 mm, 0.982 (95% CI, 0.970–0.989) for $R_{470}$ at 6.5 mm, 0.965 (95% CI, 0.943–0.982) for $R_{470}$ at 3.3 mm. The interobserver ICC for repeatability was 0.919 (95% CI, 0.869–0.951) for $R_{630}$ at 3.3 mm, 0.881 (95% CI, 0.810–0.927) for $R_{630}$ at 6.5 mm, 0.858 (95% CI, 0.775–0.913) for $R_{575}$ at 7.2 mm, 0.807 (95% CI, 0.698–0.880) for $R_{575}$ at 4.0 mm, 0.922 (95% CI, 0.874–0.953) for $R_{470}$ at 6.5 mm, 0.898 (95% CI, 0.836–0.938) for $R_{470}$ at 3.3 mm. The relationship between mean skin reflectance values and both intra- and interobserver differences are shown in S1–S12 Figs.

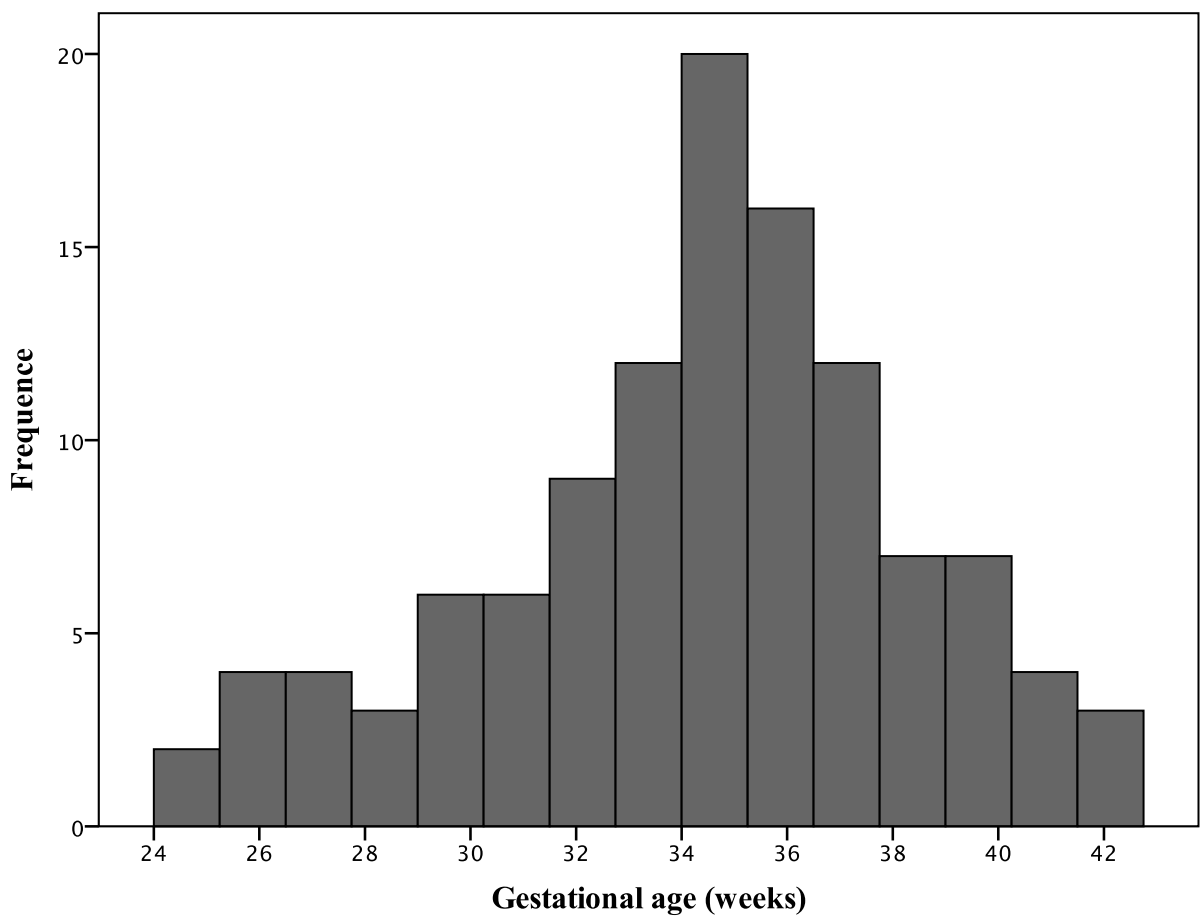

**Fig 2. Gestational age distribution in the selected sample of newborns.** This was calculated by using the gold standard approach [4]. Normal distribution (Kolmogorov-Smirnov > 5%).

**Table 1. Clinical characteristics of the studied newborns (n = 115).**

| Clinical and obstetric characteristics | N | Descriptive statistics |
|---|---|---|
| Gestational age, average (SD), weeks | 115 | 34.1 (4.1)# |
| Prematurity n (%) | 115 | 90 (78.3) |
| Major malformations n (%) | 115 | 9 (7.8) |
| Twinning n (%) | 115 | 17 (14.8) |
| Male/female n (%) | 115 | 58 (50.4)/57 (49.6) |
| NICU at the first assessment n (%) | 115 | 66 (57.4) |
| Incubator staying n (%) | 115 | 64 (55.7) |
| Phototherapy at the first assessment n (%) | 115 | 12 (10.4) |
| Birth weight, average (SD), kg | 113 | 2171.4 (783.8)# |
| Small for gestational age* n (%) | 113 | 24 (21.2) |
| Large for gestational age* n (%) | 113 | 3 (2.7) |

SD: standard deviation; NICU: neonatal intensive unit care

#Normal distribution (Kolmogorov-Smirnov > 5%)

*Based on Fenton growth-chart [18]

**Table 2. Skin reflection and location of assessment at the first and second day after birth, obtained with the optical prototype.**

| LED specifications (Reflectance values/$10^6$) | Forearm Mean (SD) | | | Sole Mean (SD) | | | Forearm vs. Sole reflectance difference | |
|---|---|---|---|---|---|---|---|---|
| | Day 1 (n = 113) | Day 2 (n = 107) | Day 1 vs Day 2 P-value* | Day 1 (n = 115) | Day 2 (n = 109) | Day 1 vs Day 2 P-value* | Day 1 (n = 228) P-value** | Day 2 (n = 216) P-value** |
| $R_{630}$ at 6.5 mm | 0.085 (0.033) | 0.078 (0.031) | **0.003**[#] | 0.108 (0.030) | 0.094 (0.028) | **< 0.001**[#] | **< 0.001**[#] | **0.001**[#] |
| $R_{630}$ at 3.3 mm | 0.444 (0.150) | 0.434 (0.144) | 0.312 | 0.507 (0.092) | 0.497 (0.099) | 0.286 | **0.001**[#] | **0.001**[#] |
| $R_{575}$ at 7.2 mm | 0.001 (0.002) | 0.001 (0.001) | 0.108 | 0.002 (0.002) | 0.001 (0.001) | 0.071 | 0.256 | 0.095 |
| $R_{575}$ at 4.0 mm | 0.004 (0.002) | 0.003 (0.001) | **0.025**[#] | 0.004 (0.003) | 0.004 (0.002) | **< 0.001**[#] | **0.004**[#] | **0.025**[#] |
| $R_{470}$ at 6.5 mm | 0.027 (0.039) | 0.025 (0.037) | 0.734 | 0.026 (0.007) | 0.032 (0.042) | 0.104 | 0.833 | 0.167 |
| $R_{470}$ at 3.3 mm | 0.202 (0.163) | 0.249 (0.036) | **< 0.001**[#] | 0.0247 (0.089) | 0.023 (0.079) | **0.012**[#] | **0.011**[#] | **< 0.001**[#] |

LED: Light emitting diode; $R_{630}$: skin reflectance at 630 nm; $R_{575}$: skin reflectance at 575 nm; $R_{470}$: skin reflectance at 470 nm;

*P-value: Paired Student t-Test;

**P-value: Independent Student t-Test;

[#]Significant association.

### Gestational age prediction based on the newborn skin reflection

The skin reflectance in the red wavelength ($R_{630}$) range was depending on GA. The univariate linear analysis showed a moderate correlation between skin reflection and GA in the skin over the sole and forearm, assessed during the first day of life, in the total group of newborns. The reflectance from LEDs at 3.3 mm/6.5 mm ratio was best associated with the weeks of gestation at the skin of the sole: r = 0.682, P < 0.001 and r = 0.710, P < 0.001, for the premature infants and the overall group of newborns, respectively (Table 3). The skin reflectance in the green 575 nm wavelength ($R_{575}$) had no significant linear correlation with the GA (Table 3).

The best-adjusted univariate model to explain the correlation between the sole skin $R_{630}$ reflection ratio (3.3 mm/6.5 mm ratio, considering distance of sensor) and the GA was an inverse function equation: $GA = 44.75 - \frac{48.93}{R_{630}\frac{3.3}{6.5}mm\ ratio}$ $R^2 = 0.56$, P < 0.001. Predicted values versus GA estimated by obstetric ultrasound had moderate correlation, R = 0.71, P < 0.001 (Fig 3). The assessment occurred at the skin of the sole, during the first day of life. The residual values of GA that were not explained by the univariate model had a normal distribution with an estimated error of 18.5 days of gestation, based on the residual values. It means that 95% of occasions, GA calculated by the model differed until 36.3 days from GA estimated by obstetric ultrasound. The values were plotted with the mean and two SD in Fig 4.

### Performance of models to predict GA based on the newborn skin reflection, including clinical variables adjustments

The skin reflection at the sole skin, $R_{630}$ 3.3 mm/6.5 mm ratio, considering distance of sensor, was influenced by clinical variables. For phototherapy, mean backscattering had values 4.16 ± 1.05 and 5.05 ± 1.18 for receiving phototherapy and no phototherapy, respectively (P = 0.023) and for incubator staying, mean backscattering was 5.58 ± 0.92 and 4.40 ± 1.13, P < 0.001 for staying inside the incubator or not, respectively. Regarding sex, mean backscattering was not different for male and female: 5.06 ± 1.21 and 4.73 ± 1.16, P = 0.136.

Compositions with the skin backscattering to $R_{470}$ and $R_{630}$, and ratios within the same colors, from the sole location on the newborn skin, were proposed as independent variables for the multivariable analysis. Table 4 introduces the final model parameters. In this selected group of neonates, 66.4% of the variability of GA was explained based on the skin reflectance.

**Table 3.** Univariate analysis of the correlation between the skin reflection and gestational age at the first day after birth, obtained with the optical prototype.

| Predictors | N | Premature infants Linear coefficient (P-value*) n = 90 | Total newborns Linear coefficient (P-value*) n = 115 |
|---|---|---|---|
| **Forearm** | | | |
| $R_{630}$ at 6.5 mm | 113 | -0.489 (< 0.001)[#] | -0.006 (0.952)[#] |
| $R_{630}$ at 3.3 mm | 113 | -0.0283 (0.795) | 0.433 (< 0.001)[#] |
| $R_{630}$ 3.3 mm/6.5 mm ratio | 113 | 0.599 (< 0.001)[#] | 0.505 (< 0.001)[#] |
| $R_{575}$ at 7.2 mm | 113 | -0.148 (0.167) | -0.086 (0.365) |
| $R_{575}$ at 4.0 mm | 113 | -0.123 (0.254) | -0.095 (0.3165) |
| $R_{575}$ 4.0 mm/7.2 mm ratio | 113 | 0.244 (0.022) [#] | -0.071 (0.457) |
| $R_{470}$ at 6.5 mm | 113 | 0.041 (0.706) | 0.076 (0.457) |
| $R_{470}$ at 3.3 mm | 113 | 0.110 (0.307) | 0.239 (0.011)[#] |
| $R_{470}$ 3.3 mm/6.5 mm ratio | 113 | 0.223 (0.037) [#] | 0.256 (0.006)[#] |
| **Sole** | | | |
| $R_{630}$ at 6.5 mm | 115 | -0.457 (< 0.001)[#] | -0.405 (<0.001)[#] |
| $R_{630}$ at 3.3 mm | 115 | 0.355 (0.001) [#] | 0.500 (< 0.001)[#] |
| $R_{630}$ 3.3 mm/6.5 mm ratio | 115 | 0.682 (< 0.001)[#] | 0.710 (< 0.001)[#] |
| $R_{575}$ at 7.2 mm | 115 | -0.260 (0.013) [#] | -0.099 (0.294) |
| $R_{575}$ at 4.0 mm | 115 | -0.178 (0.094) | -0.080 (0.396) |
| $R_{575}$ 4.0 mm/7.2 mm ratio | 115 | 0.265 (0.012) | 0.017 (0.856) |
| $R_{470}$ at 6.5 mm | 115 | 0.027 (0.799) | 0.233 (0.012) [#] |
| $R_{470}$ at 3.3 mm | 115 | 0.402 (< 0.001)[#] | 0.542 (< 0.001)[#] |
| $R_{470}$ 3.3 mm/6.5 mm ratio | 115 | 0.441 (< 0.001)[#] | 0.435 (< 0.001)[#] |

$R_{630}$: skin reflectance at 630 nm; $R_{575}$: skin reflectance at 575 nm; $R_{470}$: skin reflectance at 470 nm;

*P-value: Pearson's correlation;

[#]Significant correlation.

Modeling multiple linear regressions, models were based on combinations of skin backscattering and clinical data. The best model for gestational age prediction achieved $R^2 = 0.828$, adjusting the $R_{630}$ 3.3 mm/6.5 mm ratio and $R_{470}$ at 3.3 mm to birth weight, incubator staying, phototherapy on the first day of life, and sex. Predicted values versus GA estimated by obstetric ultrasound had correlation R = 0.91, P < 0.001 (Fig 5). The residual analysis presented a normal distribution and estimated error of 5.8 days for the GA, based on the standard residual. It means that 95% of occasions, GA calculated by the model did not differ more than 11.4 days from GA estimated by obstetric ultrasound (Fig 6).

## Discussion

In this study, we used a new approach to estimate indirectly the GA at birth, based on the optical properties of the human skin and on knowledge about the maturity of the skin, during fetal life. To obtain data, the emitter and sensor probe developed just touched the skin, for a few seconds. An electronic platform prototype with an embedded software controller made the assessment very simple and the assessment was automated by the processor. The plan for development of this process showed that this is an affordable and noninvasive solution, with risk mitigation for newborns. In the future, the device could be used, with basic instructions for caregivers at birthing locations, without previous expertise, as is needed to perform obstetric ultrasounds or neonatal maturity scores. The lack of birth care provided by skilled health personnel is a current concern occurring in approximately 30% the world and 50% in Africa (2007–2014) [21].

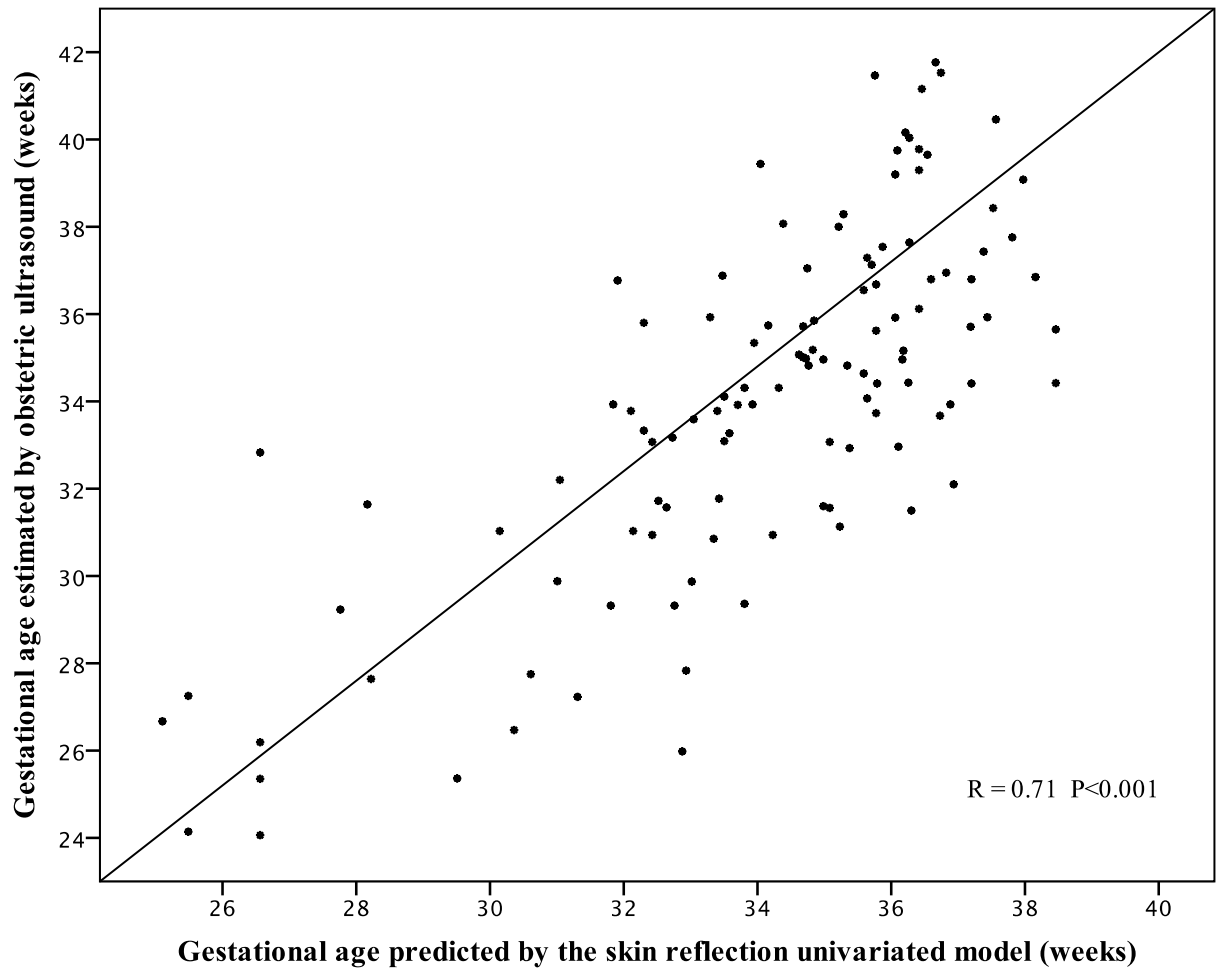

**Fig 3. Gestational age estimated by the skin reflectance, during the first day of life, at the skin of the sole of foot vs. gestational age by early obstetric ultrasound.**

## Skin reflection and GA prediction

The analysis of the neonatal skin backscattering values, in response to different colors of LEDs, showed that the optical properties of the newborn's skin can be studied at birth to aid in determining the maturity of the tissue, and secondarily calculate the gestational age. Infants that were born between 25 to 42 weeks of gestation presented increasing values of skin reflection to wavelengths of 470 nm and 630 nm. The optical device measured the backscattered light signal returned from the skin, using similar approaches to a previous report that estimated the skin fat thickness [18]. Multi-layer optical models of skin have been developed to study the effects of tissue structure on light propagation in order to determine skin thickness [22]. The skin optical properties are characterized by the absorption and scattering coefficient of the epidermis, dermis, and subcutaneous layers, according to their pigments, keratin, collagen, and blood distribution [23]. In the vascular dermis and subcutaneous layers, the main absorbers in the visible spectrum of light are hemoglobin, carotene, bilirubin, and water, while the

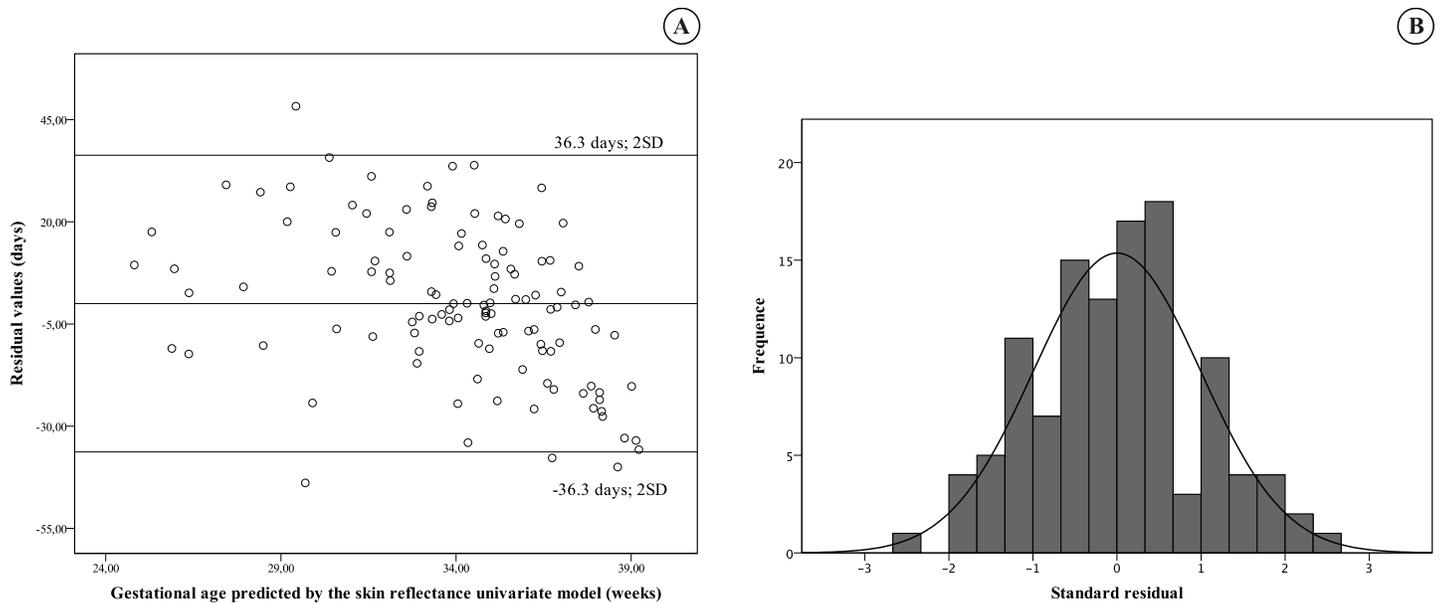

**Fig 4. Standard deviation of residual values and histogram of residual value for the skin reflectance vs. gestational age.** (A) Residual values (B) Histogram of residual values. This was during the first day of life, at the skin of the sole of the foot.

**Table 4. Predictor variables for gestational age at birth, based on the skin reflection at the sole of the newborn, adjusted for clinical variables (n = 113).**

| Predictors | Univariate Analysis Crude Correlation | | Multiple Analysis Adjusted Correlation | |
|---|---|---|---|---|
| | Beta coefficient (95% CI) | $R^2$ (P-value*) | Adjusted Beta Coefficient (95% CI) P-value | Adjusted $R^2$ (P-value*) |
| **Model 1** | | | | |
| $1/R_{630}$ 3.3 mm/6.5 mm ratio | -48.9 (-56.9 to -40.9) | 0.564 (< 0.001) | -44.4 (-53.1 to -35.8) P < 0.001 | 0.667 (< 0.001) |
| $R_{470}$ at 3.3 mm[@] | 24 (17 to 31) | 0.293 (< 0.001) | 15 (10 to 20) P< 0.001 | |
| **Model 2** | Beta coefficient (95% CI) | $R^2$ (P-value*) | Adjusted Beta Coefficient (95% CI) P-value | Adjusted $R^2$ (P-value*) |
| $1/R_{630}$ 3.3 mm/6.5 mm ratio | -48.9 (-56.9 to -40.9) | 0.564 (< 0.001) | -20.9 (-27.7 to -14.4) P < 0.001 | 0.828 (< 0.001) |
| $R_{470}$ at 3.3 mm[@] | 24 (17 to 31) | 0.293 (< 0.001) | 6 (2 to 10) P = 0.007 | |
| Birth weight (grams) | 0.004 (0.003 to 0.004) | 0.748 (< 0.001) | 0.003 (0.002 to 0.003) P < 0.001 | |
| Incubator[##] | -5.3 (-6.4 to -4.1) | 0.430 (< 0.001) | -0.77 (-1.6 to 0.1) P = 0.089 | |
| Phototherapy[##] | -4.9 (-6.2 to -1.6) | 0.089 (0.001) | Excluded (0.478) | |
| Sex[#] | -0.8 (-2.3 to 0.7) | 0.008 (0.182) | Excluded (0.552) | |

[@]value*$10^{-6}$;

[#]1 = male;

[##]1 = yes;

*Adjust of the model, ANOVA;

Constant for Model 1 = 39.5; Constant for Model 2 = 32.3.

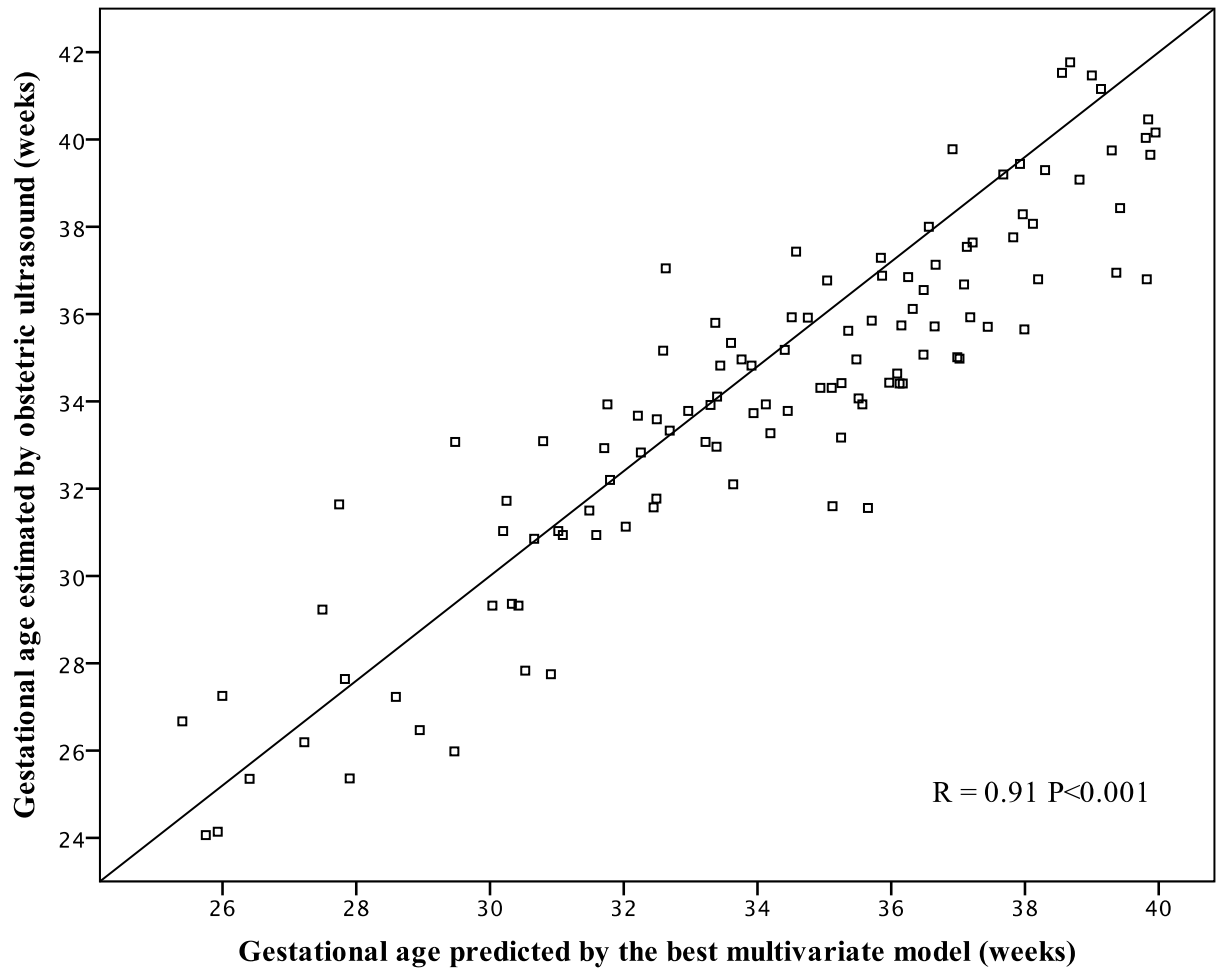

**Fig 5. Title gestational age estimated by the multivariate model vs. gestational age by the early obstetric ultrasound.**

scattering properties come from the fibrous structures, such as collagen [14]. We ascribed the lack of linear correlation between $R_{575}$ and the GA to the hemoglobin absorption properties, based on the backscattering acquisitions from the skin.

In a prior study, at 650 nm wavelength, the reduced-scattering coefficient increased linearly with gestational maturity for neonatal skin, deduced from the integrating-sphere measurements [24]. Similarly, our best correlations with the GA were at 630 nm. The $R_{630}$ at 3.3 mm/ 6.5 mm ratio showed the importance of corrections for environment influences, such as ambient light and pressure of the sensor against the skin. At 470 nm and 630 nm, the skin reflectance response changed linearly with the GA, reproducing data from other reports. One of the studies, using the 450–750 nm range, demonstrated that the total reflectance of neonatal skin, shown by the reduced-scattering coefficients, increased linearly with gestational maturity, based on the integrating-sphere approach for measurements [24]. Another study calculated reflectance at 837 nm ($R_{837}$), which was related exponentially to GA in 64 newborns at 24 to 42 weeks of gestation [17]. The authors showed that the dependence of collagen development affects the amount of optical scattering from the dermis. Despite the fact that the GA in that study was not based on the gold standard, we compared our results (cross symbol) to theirs (dot symbol), fitting our data from alive newborns together with the Lynn et al. data [17], as seen in Fig 7. The results indicated that the prototype used in our study yielded data comparable to the previous studies obtained with spectroscopes. We did not use the similar study of Post et al. to compare our data because the GA estimate was not as reliable as current methods, resulting in a gestation length of 44 weeks [16].

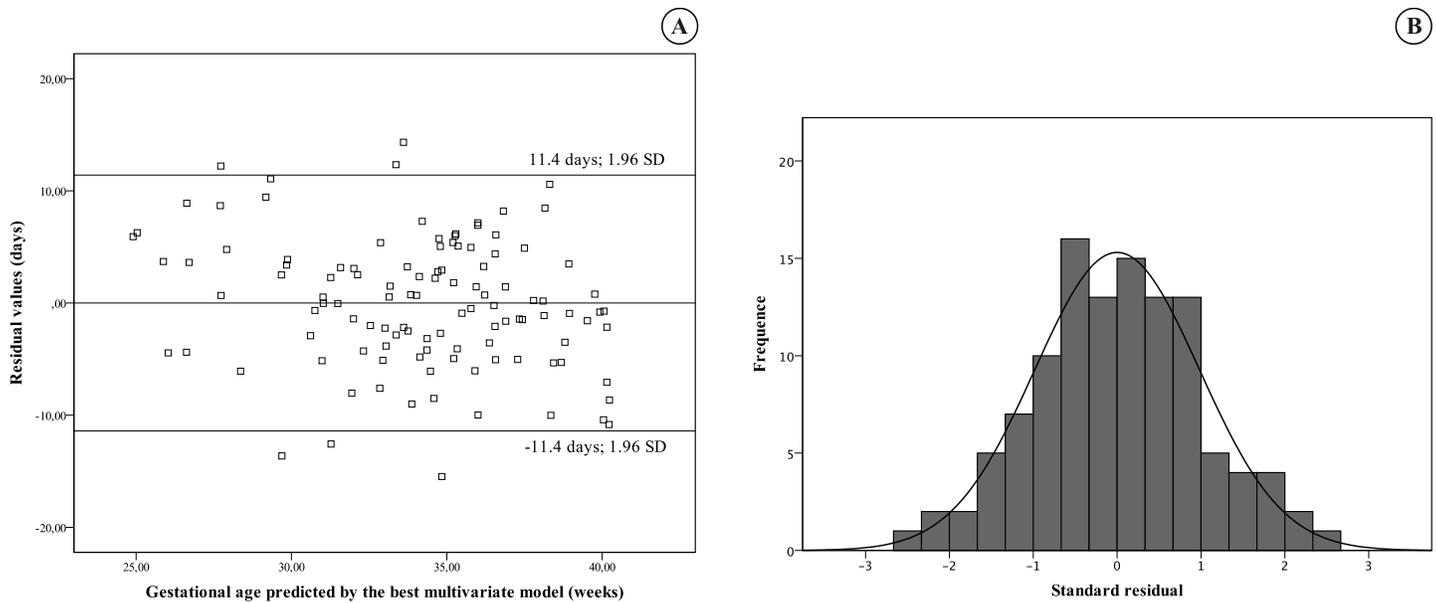

**Fig 6. Standard deviation of residual values and histogram of residual value for the best multivariate model vs. gestational age.** (A) Residual values (B) Histogram of residual values. This was during the first day of life, at the skin of the sole of the foot.

We need further analysis to understand the differences in the GA correlations between the forearm and the sole better. The neonatal skin is different from an adult in structure, function, and composition [25], mainly regarding the skin barrier function within the stratum corneum at the top of the epidermis [26]. The skin permeability barrier, formed during late gestation, is essential for neonatal survival. A report about the skin maturation in the human fetus demonstrated that this barrier formation coincides with the limit of the viability of pre-term infants 20 to 24 weeks [13], but extends until birth and through the first year of life [26]. Additionally, at the same gestational age, the skin of different places in the body, such as the abdomen and head, are in different stages of keratinization [13]. It could be an indication of new insights about the differences of optical properties between the skin of forearm and sole and non-explained outliers in our study. The skin colour and ethnicity are relevant variables to consider on the agenda for the future. It is still important to consider the topographical irregularity of the skin as a source of within-site or between-sites variations of their optical properties [27]. In the next version of our prototype, we will use synchronized and repetitive measurements at the same place to take the mean, instead of a single value, as a possible solution for the within-site variation.

## Statistical modeling choice

An equation including only the skin optical properties, $R_{630}$ and $R_{470}$ skin reflection together, explained 66.7% of the gold standard GA variability. Looking for an improved predictive model for GA, the skin reflection values were adjusted based on clinical variables.

In our sample, 78.3% of newborns were premature, 21.2% were small for gestational age, 14.8% were from twin gestation, and 7.8% had major malformations, reflecting actual settings of high complexity of care. This particular scenario is considered as part of the challenge in predicting the GA, in a health compromised population. We believe that clinical variables, easy to obtain at birthing locations, can improve the model based on the skin reflection.

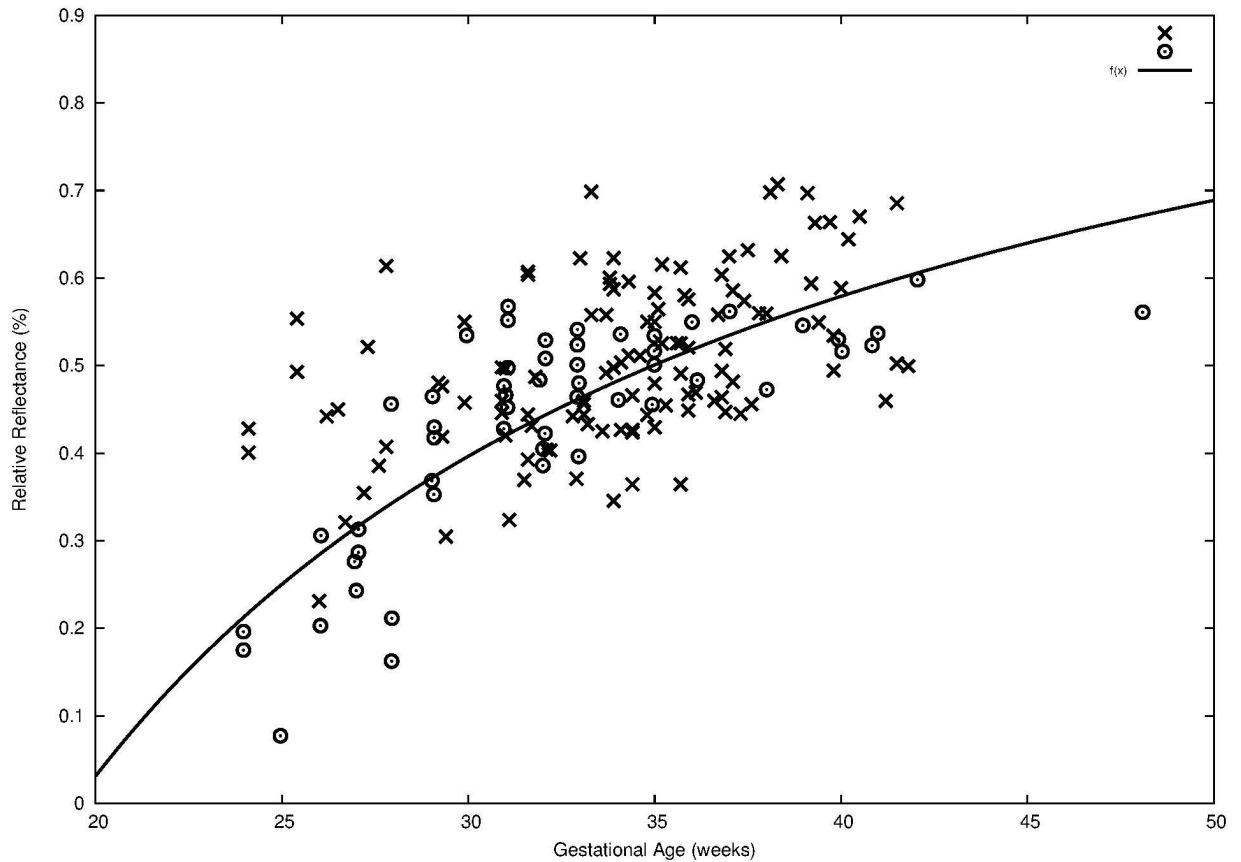

**Fig 7. The skin reflectance versus gestational age comparing different studies.** Cross symbol is the current reflectance data over the sole, at $R_{630}$ 3.3; Dot symbol represents data by Lynn at al. [17]. The line is the correspondent reflectance equation fitted on our data: $GA = 45.6 - 51.9 * R_{630}$.

Birth weight and the incubator staying variables improved the explanation regarding the GA variability, compiling weight values with $R_{630}$ and $R_{470}$ skin reflection. However, the relationship between birth weight and GA is known and this predictor alone is insufficient to estimate GA because size does not signify maturity [25]. The existence of both of the variables in the model could improve the equation, reinforcing their individual effects on the GA. Further study needs to be done on incubator staying to support or refute this initial evidence. In fact, humidity and temperature are environmental conditions associated with the optical properties of matter [28], and adjustments in the automatic calculations can be performed by the processor of the device.

Adjusted models are still important to evaluate confounding variables, such as the sex of the newborn and phototherapy, in the current analysis. The frequency of phototherapy during the first day of life was low in our studied group (10.4%). However, further studies are necessary to prove or disprove our initial results. Both size of a fetus based on ultrasound measurements, and sex, are recognized as contributing factors for standard growth [18]. This may be true for skin growth as well. The phototherapy influence on neonatal skin reflectance, as pointed out in a previous analysis [16] deserves further attention.

In terms of external validation, we expect to adjust the GA prediction model to include the challenges of identifying prematurity, in order to support health decisions, and this depends on addressing the complexity of prematurity. We understand that prematurity has a multifactorial etiology and this particular population has clinical complexity. Demographic, socioeconomic, medical, and health behavioral characteristics, as well as iatrogenic and spontaneous preterm birth, twinning are associated with prematurity, sometimes without a proper explanation [29]. Small size at birth is a risk factor associated with more than 80% of neonatal deaths and increases the risk of post-neonatal mortality [1]. A solution for predicting GA must address the inability of health systems to identify fetal growth failure, based on reliable information.

## Limitations and perspectives

A potential limitation of the present study is that reproducible measurement of CCI was possible in this first analysis, considering only two observers, accessing adult skin. Despite high coefficients of intraobserver and interobserver repeatability, three of twelve Bland–Altman plots deserve further investigation, mainly regarding the skin reflectance at 470 nm. The overall reproducibility requires confirmation in other centers and among more than two raters and preferably accessing premature and term newborns.

Parametric statistic tests are based on a theoretical probability normal distribution. The methodology was used to analyze the association between outcome and predictors, once the selected group of newborns resulted in a Gaussian distribution for the GA variable. However, in the actual birth setting, a nonparametric distribution of probability is expected for population-wide GA distribution because the prematurity rate is around 7.5% to 12.5% in the world, considering health inequities between countries [30]. Using this particular group in a cross-section design, the analysis proved the concept that newborn skin reflection is modified according to the skin growth, mirroring the gestational evolution. The main limitation of the models was generalizability. We prepared a statistical equation to explain the expected relationship between skin reflectance and GA that is not ready to be used in general population yet, but that is directed to the target population of premature infants.

Availability of reliable gestational age data is a prerequisite for preterm birth classification and health decisions [29]. Ultrasound machines and health professional training are costly, and they are not available or accessible enough to better predict GA in resource-constrained countries with fragile health care systems [6]. The skin reflection based on the new LED optical device value should be an affordable solution, but the approach needs a confirmatory study. A multiracial clinical prospective, cross-sectional multicenter study with reference standard and blinding, including low, medium, and high income countries, accepting small and large for gestational age newborns, is the next step. Additionally, artificial intelligence techniques can be utilized to improve the GA prediction.

## Supporting information

**S1 Fig. Intraobserver Bland–Altman plot for skin reflectance at 630 nm acquisition, at 3.3 mm distant of sensor.** SD Standard deviation.
(EPS)

**S2 Fig. Intraobserver Bland–Altman plot for skin reflectance at 630 nm acquisition, at 6.5 mm distant of sensor.** SD Standard deviation.
(EPS)

**S3 Fig. Intraobserver Bland–Altman plot for skin reflectance at 575 nm acquisition, at 4.0 mm distant of sensor.** SD Standard deviation.
(EPS)

**S4 Fig. Intraobserver Bland–Altman plot for skin reflectance at 575 nm acquisition, at 7.2 mm distant of sensor.** SD Standard deviation.
(EPS)

**S5 Fig. Intraobserver Bland–Altman plot for skin reflectance at 470 nm acquisition, at 3.3 mm distant of sensor.** SD Standard deviation.
(EPS)

**S6 Fig. Intraobserver Bland–Altman plot for skin reflectance at 470 nm acquisition, at 6.5 mm distant of sensor.** SD Standard deviation.
(EPS)

**S7 Fig. Interobserver Bland–Altman plot for skin reflectance at 630 nm acquisition, at 3.3 mm distant of sensor.** SD Standard deviation.
(EPS)

**S8 Fig. Interobserver Bland–Altman plot for skin reflectance at 630 nm acquisition, at 6.5 mm distant of sensor.** SD Standard deviation.
(EPS)

**S9 Fig. Interobserver Bland–Altman plot for skin reflectance at 575 nm acquisition, at 4.0 mm distant of sensor.** SD Standard deviation.
(EPS)

**S10 Fig. Interobserver Bland–Altman plot for skin reflectance at 575 nm acquisition, at 7.2 mm distant of sensor.** SD Standard deviation.
(EPS)

**S11 Fig. Interobserver Bland–Altman plot for skin reflectance at 470 nm acquisition, at 3.3 mm distant of sensor.** SD Standard deviation.
(EPS)

**S12 Fig. Interobserver Bland–Altman plot for skin reflectance at 470 nm acquisition, at 6.5 mm distant of sensor.** SD Standard deviation.
(EPS)

## Acknowledgments

The authors thank Dr. Steven L Jacques for his comments on an earlier version of the article draft, and data for the Fig 7 scatter plot: The skin reflectance at red light versus gestational age comparing different studies.

## Author Contributions

**Conceptualization:** Zilma Silveira Nogueira Reis, Rodney Nascimento Guimaraes.

**Data curation:** Zilma Silveira Nogueira Reis, Gabriela Luiza Nogueira Vitral, Ingrid Michelle Fonseca de Souza, Rodney Nascimento Guimaraes.

**Formal analysis:** Zilma Silveira Nogueira Reis, Gabriela Luiza Nogueira Vitral, Ingrid Michelle Fonseca de Souza, Rodney Nascimento Guimaraes.

**Funding acquisition:** Zilma Silveira Nogueira Reis.

**Investigation:** Gabriela Luiza Nogueira Vitral, Ingrid Michelle Fonseca de Souza, Maria Albertina Santiago Rego.

**Methodology:** Zilma Silveira Nogueira Reis, Gabriela Luiza Nogueira Vitral, Maria Albertina Santiago Rego, Rodney Nascimento Guimaraes.

**Project administration:** Zilma Silveira Nogueira Reis.

**Supervision:** Zilma Silveira Nogueira Reis, Maria Albertina Santiago Rego, Rodney Nascimento Guimaraes.

**Validation:** Gabriela Luiza Nogueira Vitral, Ingrid Michelle Fonseca de Souza, Rodney Nascimento Guimaraes.

**Writing – original draft:** Zilma Silveira Nogueira Reis, Gabriela Luiza Nogueira Vitral, Ingrid Michelle Fonseca de Souza, Maria Albertina Santiago Rego, Rodney Nascimento Guimaraes.

**Writing – review & editing:** Zilma Silveira Nogueira Reis, Rodney Nascimento Guimaraes.

## References

1. Lawn JE, Blencowe H, Oza S, You D, Lee AC, Waiswa P, et al. Every Newborn: progress, priorities, and potential beyond survival. The Lancet. 2014; 384(9938):189–205.

2. Mason E, McDougall L, Lawn JE, Gupta A, Claeson M, Pillay Y, et al. From evidence to action to deliver a healthy start for the next generation. The Lancet. 2014; 384(9941):455–67.

3. Bulletins—Obstetrics ACoP. ACOG practice bulletin. Management of preterm labor. Number 43, May 2003. Int J Gynaecol Obstet. 2003; 82(1):127–35. PMID: 12834934.

4. Committee opinion no 611: method for estimating due date. Obstet Gynecol. 2014; 124(4):863–6. https://doi.org/10.1097/01.AOG.0000454932.15177.be PMID: 25244400.

5. Wingate MS, Alexander GR, Buekens P, Vahratian A. Comparison of gestational age classifications: date of last menstrual period vs. clinical estimate. Annals of epidemiology. 2007; 17(6):425–30. https://doi.org/10.1016/j.annepidem.2007.01.035 PMID: 17395481

6. Karl S, Suen CSLW, Unger HW, Ome-Kaius M, Mola G, White L, et al. Preterm or Not–An Evaluation of Estimates of Gestational Age in a Cohort of Women from Rural Papua New Guinea. PloS one. 2015; 10 (5):e0124286. https://doi.org/10.1371/journal.pone.0124286 PMID: 25945927

7. HALL MH, CARR-HILL RA, FRASER C, CAMPBELL D, SAMPHIER ML. The extent and antecedents of uncertain gestation. BJOG: An International Journal of Obstetrics & Gynaecology. 1985; 92(5):445–51.

8. Pereira APE, Dias MAB, Bastos MH, do Carmo Leal M. Determining gestational age for public health care users in Brazil: comparison of methods and algorithm creation. BMC research notes. 2013; 6(1):1.

9. Ballard JL, Novak KK, Driver M. A simplified score for assessment of fetal maturation of newly born infants. The Journal of Pediatrics. 1979; 95(5, Part 1):769–74. http://dx.doi.org/10.1016/S0022-3476 (79)80734-9.

10. Dubowitz LMS, Dubowitz V, Goldberg C. Clinical assessment of gestational age in the newborn infant. The Journal of Pediatrics. 1970; 77(1):1–10. http://dx.doi.org/10.1016/S0022-3476(70)80038-5. PMID: 5430794

11. ERSCH J, STALLMACH T. Assessing Gestational Age From Histology of Fetal Skin: An Autopsy Study of 379 Fetuses. Obstetrics & Gynecology. 1999; 94(5, Part 1):753–7.

12. Moll I, Moll R, Franke WW. Formation of Epidermal and Dermal Merkel Cells During Human Fetal Skin Development. J Investig Dermatol. 1986; 87(6):779–87. PMID: 3782861

13. Hardman MJ, Moore L, Ferguson MW, Byrne C. Barrier formation in the human fetus is patterned. Journal of Investigative Dermatology. 1999; 113(6):1106–13. https://doi.org/10.1046/j.1523-1747.1999.00800.x PMID: 10594759

14. Lister T, Wright PA, Chappell PH. Optical properties of human skin. Journal of biomedical optics. 2012; 17(9):0909011–09090115.

15. Ho DS, Kim B-M, Hwang ID, Shin K. Evaluation of a chip LED sensor module at 770 nm for fat thickness measurement of optical tissue phantoms and human body tissue. Journal of the Korean Physical Society. 2007; 51(5):1663–7.

16. Post PW, Krauss AN, Waldman S, Auld PA. Skin reflectance of newborn infants from 25 to 44 weeks gestational age. Human biology. 1976:541–57. PMID: 976976

17. Lynn CJ, Saidi Is Fau—Oelberg DG, Oelberg Dg Fau—Jacques SL, Jacques SL. Gestational age correlates with skin reflectance in newborn infants of 24–42 weeks gestation. Biol Neonate. 1993; 64(2–3):69–75. PMID: 8260548

18. Fenton TR, Kim JH. A systematic review and meta-analysis to revise the Fenton growth chart for preterm infants. BMC pediatrics. 2013; 13(1):1.

19. Reis ZSN, Guimarães RN, inventorsDispositivo para determinação da idade gestacional, processos e usos. Brazil2016 Nov/1/2016.

20. Altman DG, Bland JM. Measurement in medicine: the analysis of method comparison studies. The statistician. 1983:307–317.

21. Organization WH. World health statistics 2015. Luxembourg: WHO Library Cataloguing-in-Publication Data; 2015. http://www.who.int/gho/publications/world_health_statistics/EN_WHS2015_TOC.pdf.

22. Yudovsky D, Pilon L. Rapid and accurate estimation of blood saturation, melanin content, and epidermis thickness from spectral diffuse reflectance. Applied optics. 2010; 49(10):1707–19. https://doi.org/10.1364/AO.49.001707 PMID: 20357850

23. Bashkatov A, Genina E, Kochubey V, Tuchin V. Optical properties of human skin, subcutaneous and mucous tissues in the wavelength range from 400 to 2000 nm. Journal of Physics D: Applied Physics. 2005; 38(15):2543.

24. Saidi IS, Jacques SL, Tittel FK. Mie and Rayleigh modeling of visible-light scattering in neonatal skin. Applied optics. 1995; 34(31):7410–8. https://doi.org/10.1364/AO.34.007410 PMID: 21060615

25. Coolen NA, Schouten KC, Middelkoop E, Ulrich MM. Comparison between human fetal and adult skin. Archives of dermatological research. 2010; 302(1):47–55. https://doi.org/10.1007/s00403-009-0989-8 PMID: 19701759

26. Telofski LS, Morello AP, Mack Correa MC, Stamatas GN. The infant skin barrier: can we preserve, protect, and enhance the barrier? Dermatology research and practice. 2012; 2012.

27. RoBERTSoN K, Rees JL. Variation in epidermal morphology in human skin at different body sites as measured by reflectance confocal microscopy. Acta dermato-venereologica. 2010; 90(4):368–73. https://doi.org/10.2340/00015555-0875 PMID: 20574601

28. Friehe CA, La Rue J, Champagne F, Gibson C, Dreyer G. Effects of temperature and humidity fluctuations on the optical refractive index in the marine boundary layer. JOSA. 1975; 65(12):1502–11.

29. Beydoun H, Ugwu B, Oehninger S. Assisted reproduction for the validation of gestational age assessment methods. Reproductive biomedicine online. 2011; 22(4):321–6. https://doi.org/10.1016/j.rbmo.2010.12.003 PMID: 21316308

30. Beck S, Wojdyla D, Say L, Betran AP, Merialdi M, Requejo JH, et al. The worldwide incidence of preterm birth: a systematic review of maternal mortality and morbidity. Bulletin of the World Health Organization. 2010; 88(1):31–8. https://doi.org/10.2471/BLT.08.062554 PMID: 20428351